\documentclass[useAMS,usenatbib,usegraphicx]{mn2e}

\usepackage{amssymb}

\newcommand{\var}{\mathop{\rm Var}\nolimits}
\newcommand{\trace}{\mathop{\rm Tr}\nolimits}
\newcommand{\Id}{\mathbfss{I}}
\newcommand{\expect}{\mathbb{E}}
\newcommand{\const}{\mathop{\rm const}\nolimits}

\title[Strategies of observations in planet searches]
{Optimal strategies of radial velocity observations in planet search surveys}
\author[R.~V. Baluev]{Roman V. Baluev\thanks{E-mail:roman@astro.spbu.ru}\\
Sobolev Astronomical Institute, St Petersburg State University, Universitetskij prospekt 28, Petrodvorets, St Petersburg 198504, Russia}
\begin{document}

\date{Accepted 2008 June 26.
      Received 2008 June 26;
      in original form 2008 May 3.}

\pagerange{\pageref{firstpage}--\pageref{lastpage}} \pubyear{2008}

\maketitle

\label{firstpage}

\begin{abstract}
Applications of the theory of optimal design of experiments to radial velocity planet
search surveys are considered. Different optimality criteria are discussed, basing on the
Fisher, Shannon, and Kullback-Leibler informations. Algorithms of optimal scheduling of RV
observations for two important practical problems are considered. The first problem is
finding the time for future observations to yield the maximum improvement of the precision
of exoplanetary orbital parameters and masses. The second problem is finding the most
favourable time for distinguishing alternative orbital fits (the scheduling of
discriminating observations).

These methods of optimal planning are demonstrated to be potentially efficient for
multi-planet extrasolar systems, in particular for resonant ones. In these cases, the
optimal dates of observations are often concentrated in quite narrow time segments.
\end{abstract}

\begin{keywords}
methods: statistical - stars: planetary systems - surveys
\end{keywords}

\section{Introduction}
\label{sec_intro}
Since the discovery of the first extrasolar planet orbiting the main-sequence star
51~Pegasi \citep{MayorQueloz95}, more than $300$ planets orbiting other stars were found,
and about $30$ extrasolar systems containing at least two planets are known (see
\emph{The Extrasolar Planets Encyclopaedia} by J.~Schneider, {\tt www.exoplanet.eu}). The
majority of these planets was discovered using the radial velocity (hereafter RV) method.
Many multi-planet systems show interesting dynamical behaviour like mean-motion resonances
and apsidal corotation locks. The sharp understanding of dynamical regimes in these
systems is of a great importance. For example, this information may provide significant
constraints on the planetary migration processes
\citep{Beauge06}.

Unfortunately, orbital parameters and masses of many of the extrasolar planets are still
poorly known due to a lack and/or a non-uniform sampling of RV observations. Estimations
of parameters may be too uncertain or alternative models of the systems may be allowed.
Unfortunately, such uncertainties are especially inherent in multi-planetary
configurations. It seems that strictly justified strategies of allocating observations
still are not widely used in the current practice of RV planet searches. The aim of the
present work is to describe several mathematically strict ways of optimal planning of RV
observations, based on the general theory of optimal design of experiments
\citep[e.g.,][chapter~10]{Ermakov,ErmZhig,Bard}, and with particular attention to refining
orbits in multi-planet configurations. In Section~\ref{sec_criteria}, several criteria of
observing schedule optimality are discussed. In Section~\ref{sec_planning}, algorithms of
optimal planning of RV observations are described in details. In Section~\ref{sec_apply},
the region of their applicability is discussed. In Section~\ref{sec_example}, the
efficiency of these algorithms is demonstrated using data for real planetary systems.

\section{Overview of optimality creteria}
\label{sec_criteria}
Let us denote $N$ RV measurements of a given star by $v_{{\rm meas},n}$, the variances of
the corresponding RV errors by $\sigma_{{\rm meas}, n}^2$. We assume that the RV errors
are statistically independent and follow Gaussian distributions with zero means and with
specified variances. The timings of the observations are denoted by $t_n$. We also assume
that the RV model (corresponding to a certain orbital configuration of the planetary
system orbiting the star) is given by $\mu(t,\btheta)$, where $\btheta$ is the vector of
$d$ parameters $\btheta$ of the RV curve for a given star (the majority of these
parameters characterises the planetary system orbiting this star). Based on the RV time
series $(t_n,v_{{\rm meas},n},\sigma_{{\rm meas},n})$ and on the model $\mu(t,\btheta)$,
we can construct an estimation $\btheta^*$ of the vector $\btheta$. Since the observations
suffer from random errors, the estimations $\btheta^*$ contains random errors as well. New
observations would decrease them. Our task is to find some `optimal' schedule for these
new observations, which would lead to a maximum improvement in the precision of the
estimations. But firstly we should clarify the sense in which we consider a given strategy
of sheduling of the observations as `optimal'. The treatment of the optimality is not
unique.

Let us write down the elements of the $d\times d$ Fisher information matrix associated
with the parameters $\btheta$:
\begin{equation}
 Q_{ij}(\btheta) = \sum_{n=1}^N \frac{1}{\sigma_{{\rm meas}, n}^2} \left.
          \frac{\partial \mu}{\partial \theta_i} \frac{\partial \mu}{\partial \theta_j}
          \right|_{t=t_n}.
\label{matrQ}
\end{equation}
The Fisher information matrix characterises the degree of statistical determinability of
the parameters $\btheta$. The larger is $\mathbfss Q$, the larger are the derivatives
$\partial\mu/\partial\theta_i$ reflecting the sensitivity of the RV model to small shifts
of the parameters, and the smaller is the domain of uncertainty associated with $\btheta$.
For example, it is well-known\citep[\S~6.4]{Lehman-est} that if the estimation $\btheta^*$
is a (non-linear) least-squares estimation and the number of observations $N$ grows
infinitely, then $\btheta^*$ is asymptotically unbiased (i.e., its mathematical
expectation $\expect\btheta^*$ tends to the true values of $\btheta$), and its probability
distribution tends to a multivariate Gaussian one with the variance-covariance matrix
$\var\btheta^*$ tending to the matrix $\mathbfss C = \mathbfss Q^{-1}$ (that is, in the
coordinate notation, $(\var\btheta^*)_{ij}\equiv \expect(\theta^*_i\theta^*_j) -
\expect\theta^*_i \expect\theta^*_j\simeq C_{ij}$). When $N$ is smaller, the distribution
of the least-squares estimations may be significantly non-Gaussian. For example, it may be
multimodal so that the data may allow several alternative orbital fits. In this case, the
variance-covariance matrix of $\btheta^*$ does not characterise the confidence domain of
$\btheta$ in advance, but the Fisher information matrix $\mathbfss Q(\btheta)$ still can
be used to characterise the degree of `peakyness' of the distribution of $\btheta^*$ in
the given point $\btheta$.

Therefore, the information matrix $\mathbfss Q$ constitutes the basis for a series of
optimality criteria. For arbitrary timings $\tau_1,\tau_2,\ldots,\tau_m$ for $m$ new
observations, we can predict the new information matrix $\tilde{\mathbfss Q}$ using the
formula similar to~(\ref{matrQ}), but containing $m$ extra terms. To obtain an optimal
schedule, we should maximize the elements of this matrix over $\tau_i$. Different criteria
from this group pay different attention to different elements of the information matrix
(see, e.g., \S2.1 by \citealt{Ermakov} and \S3.1 by \citealt{ErmZhig}). Let us select a
few popular criteria of optimality that may suit our needs:
\begin{enumerate}
\item \emph{$D$-optimality.} This criterion considers the determinant
$\det\tilde{\mathbfss Q}$ (the $D$-information) as an objective function to be maximized.
In the large-sample asymptotics ($N\to\infty$), the quantity $\sqrt{\det\mathbfss Q} =
1/\sqrt{\det\mathbfss C}$ is inversely proportional to the volume of the uncertainty
ellipsoid associated with the estimations $\btheta^*$. Therefore, the $D$-optimal schedule
lead to a maximum decrease of this volume (provided the scheduled observations are
performed).

\item \emph{Generalised $D$-optimality ($D_s$-optimality).} This criterion is used to
refine the precision of certain function of the parameters $\btheta$. Introducing the
vector of $s$ quantities $\boldeta=\boldeta(\btheta)$ to be refined, we can calculate the
matrix
\begin{equation}
 \mathbfss K = \frac{\partial \boldeta}{\partial \btheta} \mathbfss C \left(
 \frac{\partial \boldeta}{\partial \btheta} \right)^{\mathrm T},
\label{matrK}
\end{equation}
which approximates asymptotically (for $N\to\infty$) the variance-covariance matrix of
$\boldeta(\btheta^*)$. We can also predict the new matrix $\tilde\mathbfss K$ for the time
after making the new observations, by means of substituting in the eq.~(\ref{matrK}),
instead of the matrix $\mathbfss C$, the matrix $\tilde\mathbfss C=\tilde\mathbfss
Q^{-1}$. The generalised $D$-optimality criterion seeks to minimize $\det\tilde{\mathbfss
K}$.

\item \emph{$L$-optimality (linear optimality).} This criterion seeks to minimize certain
linear combination of the elements of $\tilde{\mathbfss C}$. Mathematically, we need to
minimize the quantity $\trace(\mathbfss L \tilde{\mathbfss C}) = \sum_{i,j=1}^d L_{ij}
\tilde C_{ij}$, where the positive definite or semi-definite matrix $\mathbfss L$ is fixed a
priori. This criterion may be used, for instance, to minimize a weighted average of the
variances of the estimations. The criterion of $L$-optimality minimizes the mathematical
expectation of the quadratic loss function $\sum_{i,j=1}^d L_{ij} \Delta\theta_i
\Delta\theta_j$, associated with the random errors $\Delta\btheta$ of the estimations
$\btheta^*$.
\end{enumerate}

There is some obstacle in the use of the criteria of optimality discussed above. It comes
from the fact that all of these criteria depend on the values of the parameters $\btheta$,
which are unknown. To overcome this obstacle, one of the following approaches can be used
\citep[\S\S5.2,5.3]{ErmZhig}:
\begin{enumerate}
\item \emph{Sequential approach.} In this approach, we simply substitute the current
estimations $\btheta^*$. Therefore, after obtaining more and more measurements, we refine
the estimations together with the optimality criterion.

\item \emph{Bayesian approach.} In this approach, we average the adopted objective
function using some weight function of $\btheta$. This weight function may represent the
current prior (posterior with respect to obserevations which are already made)
distribution of $\btheta$. The averaged objective function can be further maximized
(minimized) to obtain the corresponding optimal schedule.

\item \emph{Minimax approach.} In this approach, we maximize (minimize) the minimum
(maximium) value of the objective function within the domain to which $\btheta$ is
supposed to belong.
\end{enumerate}
From the computational view point, the sequential approach is more efficient, since it
does not require extra integrations or maximizations. However, it requires a bigger amount
of `priming' observations, which are needed to obtain a definite and, desirably, nearly
Gaussian, starting estimation $\btheta^*$. The Bayesian approach is formally free from
this limitation, but it has two well-known disadvantages: the complexity of the appearing
integrals over $\btheta$ and the ambiguity concerning the choice of prior distributions.
The main disadvantage of the minimax approach is the presense of extra multi-dimensional
optimisation, which is easy to perform in simplest cases only.

There is another group of optimality criteria, which deal with the Shannon information
(negative Shannon entropy) $I = -\int p(\btheta) \ln p(\btheta) d\btheta$, associated with
certain probability density $p(\btheta)$. The Bayesian approach of maximizing the Shannon
information associated with the posterior distribution of $\btheta$ was considered by
\citet{Ford08}. This method also suffers from the two mentioned disadvantages of the
Bayesian algorithms. Although for some simplified RV models \citet{Ford08} proposed a way
of decreasing the amount of calculations required in extra integrations, the abilities of
the Bayesian criteria still remain limited, especially for multi-planet systems with large
numbers of free parameters.

Since the uncertainty of the estimations $\btheta^*$ decreases for larger time series, the
sequential and Bayesian approaches become asymptotically equivalent when $N\to\infty$. The
same property of the asymptotic equivalence holds true for the $D$-optimality criterion
and the Shannon information criterion. The reason for this equivalence comes from the fact
that when $N\to\infty$, the distribution of $\btheta$ tends to the multivaraiate Gaussian
one with $\var\btheta^* \simeq \mathbfss C$. The resulting Shannon information can be
uniquely expressed via the $D$-information, according to the relation $I =
\ln\sqrt{\det\mathbfss Q} + \const$. Therefore, for relatively large $N$, the best choice
is to use the sequential optimality criterion for refining the orbital configurations of
planetary systems.

In RV planet searches, we often deal with two or more roughly equally likely orbital
solutions for the planetary system. The optimality criteria discussed above are
insensitive to this multiplicity. They favour to futher increasing of the `peakyness' of
the modes of the $\chi^2$ function or of the likelihood function (or of the posterior
distribution of $\btheta$), but they may be less useful for discriminating one of the
peaks. In the case of multiple alternative orbital fits, we should use anouther optimality
criterion, which is based on the Kullback-Leibler discriminating information (see \S5.5 by
\citealt{ErmZhig} and \S10.5 by \citealt{Bard}). Before we describe this criterion, let us
introduce some extra definitions. Given the current estimations $\btheta = \btheta^*$, we
can make a prediction of the RV at any time $t$ as $v = \mu(t,\btheta^*)$. Since
$\btheta^*$ incorporate random errors, the RV prediction should also contain a random
component leading to an uncertainty of the predicted value of $v$. For each orbital fit we
may construct its own RV prediction. After that, we can define the Kullback-Leibler
informations
\begin{equation}
I_{2|1} = \int p_1(v) \ln\frac{p_1(v)}{p_2(v)} dv, \quad
I_{1|2} = \int p_2(v) \ln\frac{p_2(v)}{p_1(v)} dv.
\label{KLinfo}
\end{equation}
Here, the probability densities $p_{1,2}(v)$ describe the distribution of the RV
prediction for the one of two orbital fits. It is not hard to see that the
quantities~(\ref{KLinfo}) represent the mathematical expectations of the likelihood ratio
statistic considering the first or the second model as true. To obtain optimal dates for
discriminating observations, we need to maximize $I_{1|2}$, $I_{2|1}$, or some their
combination.

In this paper, we aim to describe rules of optimal planning of RV measurements for
multi-planet extrasolar systems. For such systems, the initial amounts of observations,
needed in the sequential approach, are usually available. Thus we adopt below the
sequential approach to consruct detailed algorithms for $D$- and $L$-optimal scheduling
and for optimal scheduling of discriminating observations.

\section{Algorithms of optimal scheduling}
\label{sec_planning}
\subsection{D-optimal scheduling}
If we obtain an extra, $(N+1)^{\rm th}$, RV observation at time $\tau$, the precision of
the estimations $\btheta^*$ increases. The new information matrix $\tilde\mathbfss Q$ is
defined by the formula similar to~(\ref{matrQ}), but containing an extra, $(N+1)^{\rm
th}$, term in the summation. The $D$-optimality criterion seeks to find $\tau$ that
provides the maximum value of $\det\tilde\mathbfss Q$. Let us write down the asymptotic
($N\to\infty$) variance of the RV prediction $v$, using the following analogue of the
formula~(\ref{matrK}):
\begin{equation}
 \sigma_{\rm pred}^2(t,\btheta)
  = \frac{\partial \mu}{\partial\btheta} \mathbfss C \left(\frac{\partial \mu}{\partial
  \btheta}\right)^{\rm T} = \sum_{i,j=1}^d C_{ij} \frac{\partial \mu}{\partial
  \theta_i} \frac{\partial \mu}{\partial \theta_j}.
\label{pred-var}
\end{equation}
According to \citep[\S~10.3]{Bard}, the maximization of $\det\tilde\mathbfss Q$ is
equivalent to the maximization of the RV prediction variance $\sigma_{\rm pred}^2(\tau)$,
which should be calculated in the approximation~(\ref{pred-var}). That is, the optimal
time for the future RV measurement corresponds to the most uncertain RV prediction. This
rule is quite clear: to improve our knowledge, we need to make observations when our
predicting abilities are mostly limited. On contrary, we profit little from observations
made when the observed quantity is well predictable.

When merging data from several observatories, it may be more useful to maximize the full
variance of the deviation of the future measurement from the RV prediction:
\begin{equation}
 \sigma^2(\tau) = \sigma_{\rm pred}^2(\tau) + \sigma_{\rm meas}^2,
\label{full-var}
\end{equation}
where $\sigma_{\rm meas}^2$ is the variance of the future RV measurement (expected for a
given observatory). At last, we define the non-dimensional function $J^2(\tau) =
\det\tilde\mathbfss Q/\det\mathbfss Q$, which can be transformed to a more simple form
\begin{equation}
 J^2(\tau) \equiv \frac{\det\tilde\mathbfss Q}{\det\mathbfss Q} =
 \frac{\sigma^2(\tau)}{\sigma_{\rm meas}^2} = 1 + \frac{\sigma_{\rm             
 pred}^2(\tau)}{\sigma_{\rm meas}^2} \geq 1.
\label{J-simple}
\end{equation}
This relation is derived in \citep[\S10.3]{Bard} and in the Appendix~\ref{sec_algebra} of
the present paper. The identity~(\ref{J-simple}) means that the value of $J(\tau)$ tells
us how much the volume of the uncertainty ellipsoid associated with $\btheta^*$ would
decrease after making the extra observation at time $\tau$.

Often, it is not necessary to refine the whole set of parameters $\btheta$. Instead, we
may be interested in improving the precision of only some of these parameters or in
refining a certain function of $\btheta$. For instance, we have no direct need in refining
the estimation of the velocity of the barycentre of the planetary system (the constant
velocity term in the RV model). We may want to refine orbital elements of only certain
planets in the system. We may want to refine only some combinations of the parameters of
the system.

Let us assume that we need to improve the precision of the vector $\boldeta =
\boldeta(\btheta)$. Linear (asymptotic $N\to\infty$) approximation to the
variance-covariance matrix of $\boldeta$ is given by~(\ref{matrK}) The matrix
$\tilde\mathbfss K$ can be defined using the same formula but with $\mathbfss C$ changed
by $\tilde\mathbfss C$. Now we need to minimize $\det\tilde\mathbfss K$ instead of
$\det\tilde\mathbfss C$. In the case when the matrices $\mathbfss K$ and $\tilde\mathbfss
K$ are not degenerated, we can extend the definition of $J$ according to $J^2(\tau) =
\det\mathbfss K / \det\tilde\mathbfss K$. As it is shown in Appendix~\ref{sec_algebra},
the function $J(\tau)$ can be again rewritten in a more simple form:
\begin{equation}
 J^2(\tau) = \frac{\sigma^2(\tau)}{\sigma_\boldeta^2(\tau)} \geq 1,
\label{J-general}
\end{equation}
where $\sigma_\boldeta^2(\tau)$ is the conditional variance of the difference (RV
prediction $-$ actual future RV measurement), calculated under condition of fixed
$\boldeta$. This conditional variance can be calculated using the
formulae~(\ref{pred-var}) and~(\ref{full-var}), but substituting, instead of the matrix
$\mathbfss C$, the corresponding \emph{conditional} variance-covariance matrix $\mathbfss
C_\boldeta$ of $\btheta^*$:
\begin{equation}
 \mathbfss C_\boldeta = \mathbfss C - \mathbfss A^{\mathrm T} \mathbfss K^{-1} \mathbfss A
 \qquad {\rm with}\qquad \mathbfss A = \frac{\partial \boldeta}{\partial \btheta} \mathbfss C.
\label{cond-var}
\end{equation}
Here, the matrix $\mathbfss A$ represents the asymptotic approximation to the
cross-covariance matrix of $\boldeta$ and $\btheta$. Note that when $\boldeta\equiv
\btheta$, we have $\mathbfss A = \mathbfss K = \mathbfss C$, $\mathbfss C_\boldeta = 0$,
and $\sigma_\boldeta = \sigma_{\rm meas}$, as we could expect. If the parameters in
$\boldeta$ represent a subset of the parameters in $\btheta$ then the matrices $\mathbfss
K$ and $\mathbfss A$ represent certain submatrices of the matrix $\mathbfss C$ and the
calculations are much simplified. In this case, all of the elements in the matrix
$\mathbfss C_\boldeta$ are zero, except for those corresponding to the variances and
mutual correlations of the parameters $\boldeta$.

Some difficulties arise when the matrix $\mathbfss K$ is degenerated. This may take place
when some of the variables in $\boldeta$ are dependent and, hence, the matrix
$\partial\boldeta/ \partial\btheta$ is not of full rank. Such a case quite can be met in
practice and we need to process it correctly. In this case $\det\mathbfss K =
\det\tilde\mathbfss K = 0$ and $r<\dim\boldeta$ axes of the uncertainty ellipsoid of
$\boldeta$ vanish. However, the volume of this ellipsoid in the subspace of the resting
$(\dim\boldeta-r)$ axes does not vanish yet. This volume is proportional to the product of
all non-zero eigenvalues of $\mathbfss K$ (recall that the product of all eigenvalues of a
matrix is equal to its determinant). Denoting this product as $\mathcal D \mathbfss K$, we
can define $J^2(\tau) = \mathcal D\mathbfss K / \mathcal D \tilde\mathbfss K$. Note that
this revised definition incorporates the previous one as a special case. Again, the
function $J(\tau)$ can be transformed to the more simple form~(\ref{J-general}). However,
now we cannot calculate the inverse matrix $\mathbfss K^{-1}$ in the eq.~(\ref{cond-var}).
Instead, we should substitute the pseudoinverse matrix $\mathbfss K^+$. This pseudoinverse
matrix can be calculated via the eigendecomposition of $\mathbfss K$ (see, e.g.,
Appendix~A in \citep{Bard} for a brief summary of this procedure and further references).

In the general case, we need to find the time range when the values of the
function~(\ref{J-general}) are large. The physical sense of this rule is intuitively clear
again: to improve the precision of a given set of parameters, we ought to make
observations when the uncertainty of our prediction is large, in comparison with the same
uncertainty calculated under assumption that the parameters to be refined are known
exactly.

It is worth noting that the variances $\sigma_{\rm pred}^2(\tau)$, $\sigma_{\boldeta,\rm
pred}^2(\tau)$, and, hence, the values of $J(\tau)$ are invariable with respect to
arbitrary smooth non-degenerated re-parametrization. That is, if we define the
transformations of parameters $\btheta_1 = \btheta_1(\btheta)$ and $\boldeta_1 =
\boldeta_1(\boldeta)$ having non-zero Jacobians (i.e., $\det(\partial
\btheta_1/\partial\btheta)\neq 0$ and $\det(\partial \boldeta_1/\partial\boldeta)\neq 0$)
in the point $\btheta^*$, the values of $J(\tau)$ calculated by $\btheta_1,\boldeta_1$
would exactly coincide with those calculated by $\btheta,\boldeta$.

\subsection{$L$-optimal scheduling}
\label{sec_Lcrit}
We may also be interested in constructing an $L$-optimal schedule. Now the objective
function to be minimized by $\tau$ is $\trace(\mathbfss L \tilde\mathbfss C)$. We can
define the non-dimensional gain function $l(\tau) = \trace(\mathbfss L \mathbfss
C)/\trace(\mathbfss L \tilde\mathbfss C)$ to be maximized. Using identity~(\ref{..}) from
Appendix~\ref{sec_algebra}, we can write down the expression
\begin{equation}
\frac{1}{l(\tau)} = 1 - \frac{\bmath c(\tau)^{\rm T}\, \mathbfss L\, \bmath
c(\tau)}{\sigma^2(\tau)\trace(\mathbfss L \mathbfss C)},
\label{l-simple}
\end{equation}
where the vector $\bmath c(\tau)$ has elements $c_i= \sum_{j=1}^d C_{ij}
\frac{\partial\mu}{\partial\theta_j}$ and represents the asymptotic covariation of
$\btheta^*$ and the RV prediction at $t=\tau$.

When we want the refine the vector of parameters $\boldeta(\btheta)$, we need to use some
generalisation of the function~(\ref{l-simple}). In this case, we need to minimize the
function $\trace(\mathbfss L \tilde\mathbfss K)$ by $\tau$. We redefine $l(\tau) =
\trace(\mathbfss L\mathbfss K)/\trace(\mathbfss L \tilde\mathbfss K)$ and use the first of
the formulae~(\ref{...}) to write down
\begin{equation}
\frac{1}{l(\tau)} = 1 - \frac{\bmath a(\tau)^{\rm T}\, \mathbfss L\, \bmath
a(\tau)}{\sigma^2(\tau)\trace(\mathbfss L \mathbfss K)} = 1 - \frac{\bmath
c(\tau)^{\rm T} \mathbfss M \bmath c(\tau)}{\sigma^2(\tau)\trace(\mathbfss M
\mathbfss C)},
\label{l-general}
\end{equation}
where the vector $\bmath a(\tau)$ having elements $a_i = \sum_{j=1}^d A_{ij}
\frac{\partial\mu}{\partial\theta_j}$ represents the asymptotic covariation of
$\boldeta(\btheta^*)$ and the RV prediction at $t=\tau$. The matrix $\mathbfss M$ is equal
to $(\frac{\partial \boldeta}{\partial \btheta})^{\rm T} \mathbfss L (\frac{\partial
\boldeta}{\partial \btheta})$. This generalised $L$-optimality rule represents the
usual one with the matrix $\mathbfss L$ changed by $\mathbfss M$. Note that now possible
degeneracy of the matrix $\mathbfss K$ does not produce any obstacles.

The $L$-optimal criterion is invariable with respect to non-degenerated changes of
variables only if the matrix $\mathbfss L$ is transformed in accordance: $\mathbfss L_1 =
(\frac{\partial \boldeta}{\partial \boldeta_1})^{\rm T} \mathbfss L (\frac{\partial
\boldeta}{\partial \boldeta_1})$.

\subsection{Scheduling discriminating observations}
Let us now assume that we have two alternative RV models $\mu_1(t,\btheta_1)$ and
$\mu_2(t,\btheta_2)$ describing two different orbital configurations of the planetary
system. The sets of parameters, $\btheta_1$ and $\btheta_2$, do not necessarily coincide
and the dimensions of the models, $\dim \btheta_1$ and $\dim \btheta_2$, may be different
as well. Using the approach from the previous subsection, we can calculate the predictions
$v_1(\tau)$ and $v_2(\tau)$ along with the full variances $\sigma_1^2(\tau)$ and
$\sigma_2^2(\tau)$, according to~(\ref{full-var}). Then we can write down the expected
information for discriminating between the models \citep[\S10.5]{Bard}, $J_{12}(\tau)$, as
\begin{equation}
 J_{12} = - 1 + \frac{1}{2} \left( \frac{\sigma_1^2}{\sigma_2^2} +
                                   \frac{\sigma_2^2}{\sigma_1^2} \right) +
          \left(\frac{1}{\sigma_1^2} + \frac{1}{\sigma_2^2} \right) \frac{(v_1-v_2)^2}{2}.
\label{info-discrim}
\end{equation}
The function~(\ref{info-discrim}) represents the sum of the Kullback-Leibler informations
$I_{1|2}$ and $I_{2|1}$, calculated from~(\ref{KLinfo}) under assumption that the
distributions of $v_i$ are close to Gaussian. The largest values of $J_{12}(\tau)$
correspond to the most promising time for future observations to rule out one of the
alternative models. This rule means that we need to make RV observations when the two
models imply largely different predictions of the radial velocity. Simultaneously, the
uncertainties of these predictions should not be too large, in order to avoid
statistically insignificant differences. The combination~(\ref{info-discrim}) takes into
account both these requirements.

The property of the invariance of $J(\tau)$ with respect to a re-parametrization is valid
for the function $J_{12}(\tau)$ as well.

\subsection{Scheduling multiple observations}
We may need to plan several observations simultaneously. This problem may arise, for
instance, when we wish to plan (at least preliminarily) a whole set of observation
allocated for a coming observing season for a given star.

Let us denote the timings of $m$ future observations by $\tau_1,\tau_2,\ldots, \tau_m$.
Using the same approach as in the previous subsections, we can make $m$ RV predictions
forming the vector $\bmath v$ ($\dim \bmath v=m$) and calculate the full
variance-covariance matrix $\mathbfss V_{\rm pred}$ of this vector. Thus the $m\times m$
matrix $\mathbfss V_{\rm pred}$ should contain the variances and cross covariations of the
RV predictions. Also, we can calculate the conditional $m \times m$ variance-covariance
matrix $\mathbfss V_{\boldeta,\rm pred}$ of the vector $\bmath v$, taken under condition
of fixed $\boldeta$. Corresponding variance-covariance matrices of full deviations of the
future RV measurements from their predictions can be calculated as
\begin{equation}
\mathbfss V = \sigma_{\rm meas}^2 \Id + \mathbfss V_{\rm pred}, \qquad
\mathbfss V_\boldeta = \sigma_{\rm meas}^2 \Id + \mathbfss V_{\boldeta,\rm pred}
\end{equation}
with $\Id$ being the identity matrix. Now we can write down the extension of the function
$J(\tau)$ to multiple observations:
\begin{equation}
 J^2(\tau_1,\tau_2,\ldots,\tau_m) \equiv \frac{\mathcal D\mathbfss K}{\mathcal D\tilde\mathbfss K}
     = \frac{\det\mathbfss V}{\det\mathbfss V_\boldeta} \geq 1.
\label{J-mult}
\end{equation}
The generalisation of the function $l(\tau)$ can be calculated according to the equality
\begin{equation}
\frac{1}{l(\tau_1,\tau_2,\ldots,\tau_m)} \equiv \frac{\trace(\mathbfss L \tilde\mathbfss
K)}{\trace(\mathbfss L\mathbfss K)} = 1 - \frac{\trace(\mathbfss V^{-1}
\mathbfss W)}{\trace(\mathbfss L\mathbfss K)},
\end{equation}
where the matrix $\mathbfss W$ is calculated in the same way as the variance-covariance
matrix $\mathbfss V_{\rm pred}$ of RV predictions but substituting, instead of the matrix
$\mathbfss C$, the matrix $\mathbfss C\mathbfss M\mathbfss C$ (with $\mathbfss M$ given in
Section~\ref{sec_Lcrit}). The generalised discriminating information looks like
\begin{eqnarray}
 J_{12}(\tau_1,\tau_2,\ldots,\tau_m) = -m + \frac{1}{2}\trace\left( \mathbfss V_1^{-1}
     \mathbfss V_2 + \mathbfss V_2^{-1} \mathbfss V_1 \right) + \nonumber\\
   + \frac{1}{2} (\bmath v_2 - \bmath v_1)^{\rm T}
      \left( \mathbfss  V_1^{-1} + \mathbfss V_2^{-1} \right) (\bmath v_2 - \bmath v_1),
\label{info-discrim-mult}
\end{eqnarray}
where subscripts of $\mathbfss V$ and $\bmath v$ refer to the alternative models.

The main difficulty in using time allocation rules based on the
functions~(\ref{J-mult}--\ref{info-discrim-mult}) may be connected with too large number
of timings $\tau_i$ to be found. Perhaps, there is no big obstacles to scan a
two-dimensional grid of $(\tau_1,\tau_2)$ directly when $m=2$. For $m\geq 3$, such direct
scanning requires too intensive computations and is not practical. We may use here the
following algorithm. At first, the one-dimensional objective function $\Phi_1(\tau_1)$ is
constructed (it may be $J(\tau_1)$, $l(\tau_1)$, or $J_{12}(\tau_1)$, depending on our
aims). Using the direct one-dimensional search of the maximum, the date $\tau_1$ for one
of the future RV measuments is obtained. Then the two-dimensional planning function,
$\Phi_2(\tau_1,\tau_2)$, is plotted, but the date $\tau_1$ is fixed at the value obtained
in previous step. Therefore, again we can use a one-dimensional search of the maximum. As
a result, we obtain the second date $\tau_2$. The values of $\tau_1$ and $\tau_2$ may be
further adjusted using some non-linear maximization algorithm. Then we construct the
function $\Phi_3(\tau_1,\tau_2,\tau_3)$ with $\tau_1$ and $\tau_2$ fixed, obtain the
optimal value of $\tau_3$, adjust the whole array $\tau_1,\tau_2,\tau_3$, and so on until
the full set of $m$ optimal dates is found. We may break this sequence if an extra
observation does not provide enough gain. The released time can be used to observe other
stars.

\section{Applicability of the algorithms}
\label{sec_apply}
Of course, there is no statistical method that can be applied in every practical
situation. The algorithms described above require the following conditions to be
satisfied:
\begin{enumerate}
\item The estimations $\btheta^*$ are obtained using the least-squares approach or the
approach used to obtain them is verified to be equivalent in the sense of planning the
observations.

\item The number of existing observations should be sufficient, so that the estimations
$\btheta^*$ are (approximately) unbiased and their joint distribution can be approximated
by a multivariate Gaussian one.

\item The equations of the RV model, $\mu(t,\btheta)$, and of the parameters to be
refined, $\boldeta(\btheta)$, can be linearised in the uncertainty ellipsoid surrounding
the vector of the estimations.
\end{enumerate}

The first condition is usually satisfied in practice. Here, it is worth mentioning the
paper \citep{Baluev08b} where an approach, other than the least-squares one, was proposed
for determination of orbital parameters and masses of exoplanets. This maximum-likelihood
approach incorporates a built-in estimation of the so-called RV jitter to be taken into
account in the estimations of planetary parameters. A careful analysis shows that the
modification of the rules of optimal planning for this case is easy and straightforward.
This is provided by the fact that the cross elements in the Fisher information matrix,
corresponding to the estimations of $\btheta$ and of the RV jitter, vanish. In fact, the
rules of planning remain the same, but with the clause that the RV jitters should not
enter in the vector $\btheta$. Instead, they should only be added to the values of
$\sigma_{\rm meas}^2$.

The second condition is satisfied for well-conditioned situations, when there is a single
clear best-fitting orbital model of the system or the alternative models are well
separated in the parametric space. The cases when RV data allow multiple locally
best-fitting orbital solutions located suspiciously close to each other, correspond to the
ill-conditioned situation. In such cases, no one of the alternative orbital solutions may
represent a good approximation to the real configuration of the system. The local maxima
of the likelihood function (or local minima of the weighted r.m.s.) represent only
`ripples' produced by the lack of the data. The formal uncertainties of the estimations
may be unrealistic and underestimated. New RV measurements may change the orbital
solutions dramatically. In this situation, the estimations of parameters of a planetary
system are strongly biased and their distribution is far from the Gaussian one. Then the
optimal planning algorithms should be used with care. We should track the sensitivity of
their results to what orbital solution we adopt, to the functional model of the RV curve,
to the set of free parameters. Seemingly, the extra-solar systems HD82943 \citep{Mayor03a}
and HD37124 \citep{Vogt05} may represent such cases.

Different methods can be used to assess the reliability of orbital fits, needed to justify
the use of the scheduling rules described above. \citet{Beauge08} performed a series of
orbital fits of the system of HD82943 with truncated RV datasets, in order to prognose the
sensitivity of current orbital configuration to future RV measurements. Another, probably
more rapid, approach was used by \citet{Baluev08c} for the system of HD37124. It uses the
condition number of the Fisher information matrix $\mathbfss Q$ (or, speaking more
precisely, of the scaled information matrix having elements $Q_{ij} /
\sqrt{Q_{ii} Q_{jj}}$) to assess the degree of `ravineness' of the graph of the likelihood
function.

The third condition is tightly connected with the second one. For robust cases, both these
conditions should hold true asymptotically, when $N$ grows (because then all uncertainties
tend to zero). For a finite $N$, the temporal region of their validity is limited, mainly
due to the non-linear dependence on orbital periods of planets. It is admissible to use
the linear methods of optimal scheduling during the time much less than the span of the RV
time series (say, less than one third of the total time span). Attempts of prediction of
optimal dates in a more distant future may be unsafe.

Not every kind of non-linearity and non-gaussianity of the parameters can make the linear
theory of planning the observations unreliable. For instance, when the orbital
eccentricity of a planet is small, its distribution may be non-Gaussian, though the
orbital configuration is well-determined. This is a typical situation for the systems
containing a hot Jupiter planet: the true orbital eccentricity $e$ of the hot Jupiter may
be so small that even a very precise determination is unable to detect its deviation from
zero. The formal (linear) estimation may look like $e = 0.01 \pm 0.01$, implying the
argument of the periastron $\omega$ is ill-determined. In this case, we can make the
following change of variables: $x=e\cos\omega, y=e\sin\omega$. The dependence of the
radial velocity on the new pair of parameters $x,y$ is almost linear and all necessary
equations can be perfectly linearised with respect to these new variables. The joint
distribution of estimations of $x,y$ is much closer to a bivariate Gaussian (peaked near
zero). In practical calculations, it is not necessary to perform such change explicitly,
thanks to the invariance property of the functions $J(\tau)$ and $J_{12}(\tau)$. We can
calculate $J(\tau)$ and $J_{12}(\tau)$ in the common way, using linear approximations with
the initial (non-linear and non-Gaussian) set of parameters. As it was discussed above,
the result is exactly the same as for the new (almost linear and Gaussian) set.

We also note that rules described in Section~\ref{sec_planning} do not account for
possible sunlight or moonlight contamination: the formal maximum of $J(\tau)$ or
$J_{12}(\tau)$ may lie beyond the observing window of a given star. Therefore, we should
search for maximum of $J$ and $J_{12}$ within the admissible dates only.

\section{Applications}
\label{sec_example}
\subsection{Gliese 876: an orbital resonance}
\begin{figure*}
\includegraphics[width=\textwidth]{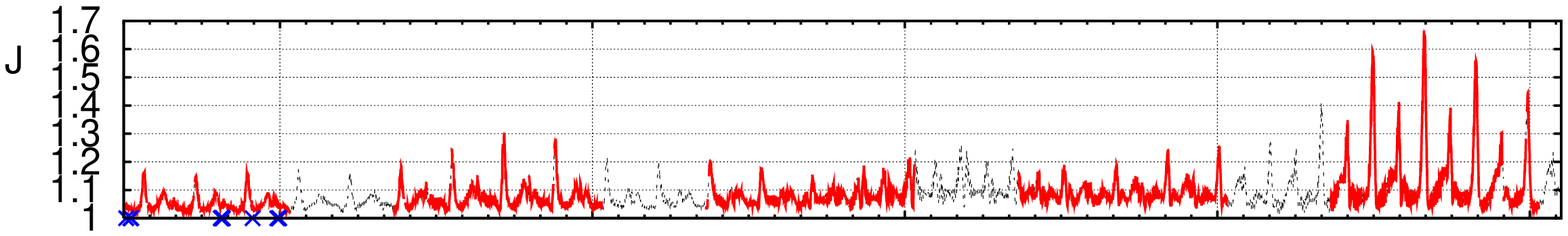}\\
\includegraphics[width=\textwidth]{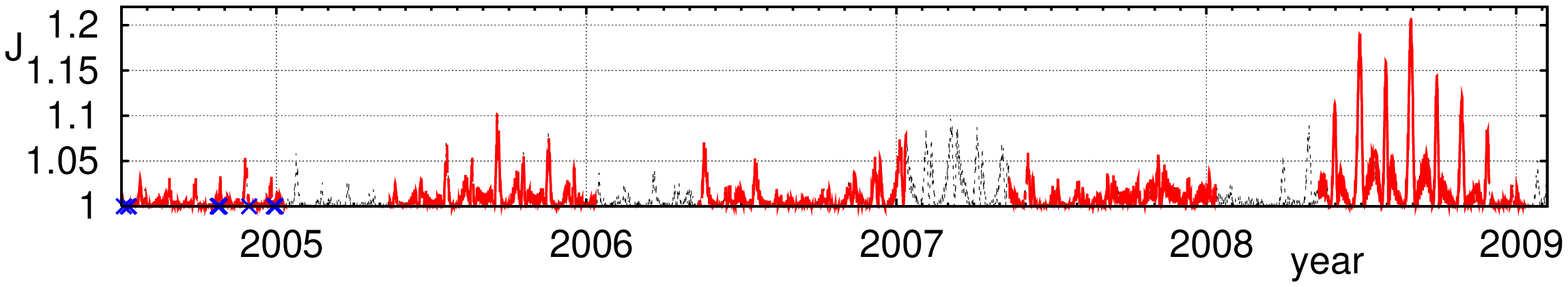}
\caption{The graphs of the function $J(\tau)$ for the planetary system of GJ876. The
broken segments of the graph correspond to the time when observations are impossible (when
the star is absent on the night sky or is contaminated by the moonlight). The crosses on
the time axis mark positions of several last RV measurements. Top panel: the vector
$\boldeta$ incorporates $10$ parameters of the planets b and c and the common orbital
inclination to the sky plane. Bottom panel: $\boldeta$ incorporates only the orbital
inclination. See text for more details.}
\label{fig_GJ876}
\end{figure*}
The first (Jovian) planet in this system was discovered by \citet{Delfosse98} using the
spectrograph ELODIE and shortly after this confirmed by \citet{Marcy98}, basing on RV
observations at Keck observatory. Some further, \citet{Marcy01} announced the second
Jovian planet, which was trapped in the 2/1 mean-motion resonance. Now this system is
believed to host three planets: the very low-mass ($\sim 7.5$ Earth mass) planet d on a
short-period ($P_d\approx 2$~days) orbit and two Jovian planets b and c, trapped in the
2/1 mean-motion resonance with $P_c\approx 30$~days and $P_b\approx 60$~days
\citep{Rivera05}. It is important that the gravitational interactions between planets b
and c were directly observed in the RV curve: after $\sim 8$~years of observations, the
orbital periastra of these planets have completed a full revolution. It makes possible to
determine the inclination of the system to the sky plane (it was estimated by about
$50^\circ$), but simultaneously it introduces extra statistical uncertainties and
correlations between different parameters. In addition, the orbital period of the planet c
is close to lunar cycles. All these facts make precise determination of orbital parameters
and masses in the system more difficult. Although currently the orbits in this system are
constrained well, we may be interested in further refining the estimations and suppressing
the correlations between different parameters of this system.

We adopt the three-planet RV model taking into account planetary perturbations. The orbits
are assumed to lie in a common plane, and its inclination to the sky plane is treated as
an extra free parameter. Therefore, the RV model have total of $17$ free parameters
$\btheta$: four osculating orbital elements and mass for the planets b, five similar
parameters for the planet c, five ones for the planet d, the orbital inclination and the
constant velocity term. The RV dataset was published in \citep{Rivera05} and consists of
$N=155$ Keck RV measurements with typical internal RV uncertainties $\sim 4$~m/s. After
obtaining the best-fitting estimations $\btheta^*$, we can apply the algorithm of the
$D$-optimal scheduling discussed above. Here and in further examples, to obtain a more
accurate estimation of the RV error variance $\sigma_{\rm meas}^2$, expected from the
observation being scheduled, we use the maximum-likelihood approach of estimating the RV
jitter, described in \citep{Baluev08b}.

In Fig.~\ref{fig_GJ876}, the function $J(\tau)$ for this system is plotted, for the case
when $\boldeta$ incorporates the $10$ parameters of planets b and c and the common orbital
inclination (case I), and for the case when $\boldeta$ incorporates only the common
orbital inclination (case II). We can see that these graphs are peaked. The sequence of
peaks shows approximately monthly periodicity, probably due to an interaction between the
RV periodicities inspired by the resonant planet `c' and `b' and the aliasing inspired by
lunar cycles. During the observing season following the last published observations, the
heights of the largest peaks corresponded to about $30\%$ decreasing of the total volume
of the uncertainty ellipsoid of $\boldeta$ (in the case I) and about $10\%$ increasing of
the accuracy of the orbital inclination (in the case II). Such increase of precision would
be provided by a single RV measurement only. It is important that these peaks are not
necessarily centered in the regions where the observations cannot be done due to the
moonlight contamination. Typically, the peaks have semi-widths of about one week only. If
made three days before or after a maximum of $J(\tau)$, the observations would not yield
much gain. We can see that, during the last season when the star was observed, some peaks
of $J(\tau)$ were not covered by observations\footnote{As it can be seen in
Figs.~\ref{fig_GJ876} and~\ref{fig_HD208487}, the actual RV observations always fall near
minima on the graphs. This fact does not mean that all these observations were sampled so
inefficiently. This means that \emph{extra} observations would yield little gain in these
positions, because they would only duplicate the existing observations.}. Possibly, the
observations might be distributed more efficiently, if this graph was constructed at that
time. Also, a series of tempting peaks of $J(\tau)$ can be seen in further seasons.
However, the probability to observe in these narrow segments is quite small until the
optimal planning algorithms are not used.

We can see that, according to the graphs in Fig.~\ref{fig_GJ876}, the most tempting time
ranges for making the RV observations in the nearest future are located in the end of
August, 2008 and in the end of October, 2008. Unfortunately, any more precise statement of
the optimal dates would be unreliable, because our analysis incorporates only published RV
measurements of $3-4$ years old, whereas their time span is only $8$ years. It may be
dangerous to predict the optimal dates which are spaced from the last real measurement by
about half of the actual time span. Moreover, it is possible that the actual RV
measurements spanning the last three observing seasons of GJ876, significantly shift the
positions of the peaks of $J(\tau)$. Nevertheless, the existence of favourable observation
dates for GJ876 in the nearest future is clear. The precise dates can be determined on the
basis of the up-to-date RV dataset, including the unpublished measurements taken in the
observing seasons of 2005-2007.

\subsection{HD208487: an alias ambiguity}
\begin{figure*}
\includegraphics[width=\textwidth]{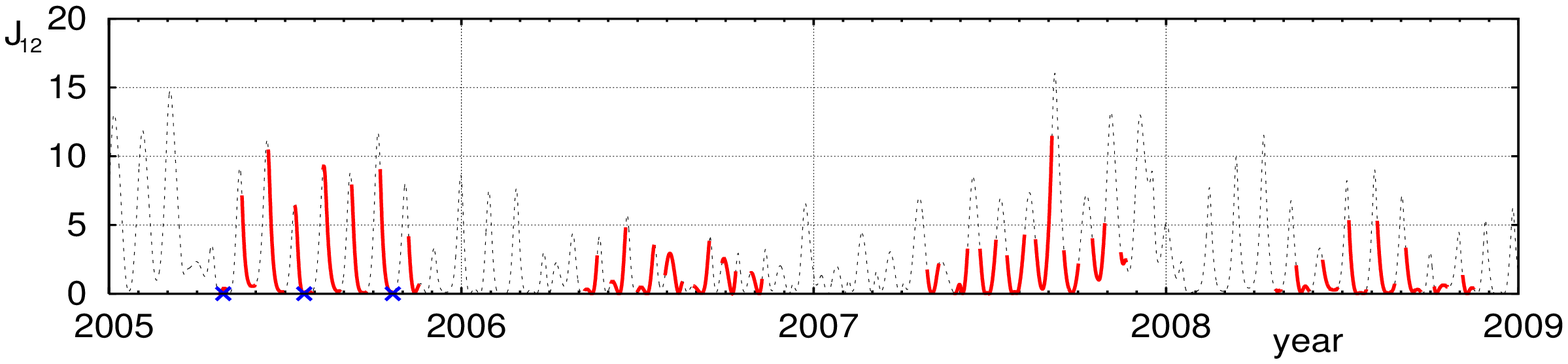}
\caption{The graph of the information for discrimination between two orbital models for
the star HD208487 described in the text. The thin broken segments of the graph correspond
to the time when observations are impossible. The crosses on the time axis mark positions
of three last RV measurements.}
\label{fig_HD208487}
\end{figure*}
The first, roughly half Jupiter mass, planet in this system was discovered by
\citet{Tinney05} on the basis of RV observations made at the Anglo-Australian Observatory.
Its orbital period is close to $130$~days. \citet{Gregory05,Gregory07} found an extra
periodicity in the RV data for this star. The orbital period of the putative second planet
was estimated by about $1000$~days, and its mass was estimated to be close to the mass of
the first planet. \citet{Wright07} found two alternative solutions, one is consistent with
that from \citet{Gregory07}, and another one corresponds to the orbital period of the
second planet of about $28.6$~days.

The periodogram of the latest RV data for HD208487 ($N=35$ measurements published by
\citealt{Butler06} and having the internal precision $\sim 5$~m/s) indeed shows two high
peaks, near $28.6$~days and near $900$~days. The difference between the respective
frequencies corresponds to a period of about $29.5$~days, clearly indicating the aliasing
connected with the full moon / new moon cycle. The false alarm probability associated with
the higher peak (which is near $28.6$~days) can be estimated using analytic bounds from
\citep{Baluev08a} by less than $1\%$. Thus, some extra periodicity is probably present,
but it is not fully clear which periodogram peak is real and which one is its alias.
Naturally, we might ask, when the new RV observations should be made for them to rule out
one of these alternatives.

In Fig.~\ref{fig_HD208487}, the corresponding graph of the function $J_{12}(\tau)$ is
plotted. It was constructed using two best-fitting double-Keplerian models of the RV
curve, with the orbital period of the putative second planet near $900$~days or near
$28.6$~days. In both models, the number of free parameters of the fit was equal to $11$
($5$ and $5$ usual parameters of the Keplerian RV variation for the two planets plus
constant velocity term). We can see that, in the last observating season in 2005, there
were many good time segments when extra observations would be highly desirable. In these
peaks of $J_{12}(\tau)$, the RV predictions for different orbital models diverged by up to
$\sim 20$~m/s. A single observation placed in one of these peaks could rule out one of the
alternative models at the significance level of $\sim 2$ sigma or even more. However,
three RV measurements actually made during this season did not cover the peaks of
$J_{12}(\tau)$. It is remarkable that the subsequent observing seasons offer less
opportunities of discrimination between the models.

\section{Conclusions}
\label{sec_conc}
This paper demonstrates how the general tools of the theory of optimal design of
experiments can be used for planning RV observations in planet search surveys. Two
important practical problems were considered. In the first one, the observations are
required to produce the largest improvement in the precision of estimations of
exoplanetary parameters. In the second one, the observations are planned to yield maximum
information for distinguishing between two alternative orbital models of an exoplanetary
system.

These optimising tools are demonstrated using RV data for several real planetary systems.
It is shown that these tools may significantly increase the efficiency of observations in
RV planet search surveys. They would be especially useful for multi-planet extrasolar
systems, in particular for systems containing planet pairs in a mean-motion resonance, for
many of which the orbits are still poorly determined. They include, among others, the
well-known systems of GJ876, HD82943, and possibly HD37124. The algorithms of optimal
scheduling may also be useful in resolving ambiguities concerning planetary orbital
configurations, e.g. the alias ambiguity for the HD208487 system.

\section*{Acknowledgments}
This work was supported by the Russian Foundation for Basic Research (Grant 06-02-16795)
and by the President Grant NSh-1323.2008.2 for the state support of leading scientific
schools. I am grateful to Profs. K.V.~Kholshevnikov and V.V.~Orlov for comments which
helped to improve this manuscript. Also, I would like to thank the anonymous referee for
providing suggestions of a great importance.

\bibliographystyle{mn2e}
\bibliography{planning}

\appendix

\section{Some matrix relations}
\label{sec_algebra}
Let us consider the matrix $\mathbfss Q$ given by~(\ref{matrQ}) and the matrix
$\tilde\mathbfss Q = \mathbfss Q + \bmath g \bmath g^{\rm T}/\sigma_{\rm meas}^2$, where
the elements of the vector $\bmath g$ are given by $g_i = \partial\mu/\partial\theta_i$.
Using the identity $\det(\Id + \mathbfss A \mathbfss B) = \det(\Id + \mathbfss B \mathbfss
A)$ from \citep[Appendix~A]{Bard}, which is valid for arbitrary matrices $\mathbfss A$ and
$\mathbfss B$ with matching dimensions, we can obtain the relation
\begin{equation}
\frac{\det\tilde\mathbfss Q}{\det\mathbfss Q} = \det\left(\Id + \frac{\mathbfss Q^{-1}
(\bmath g \bmath g^{\rm T})}{\sigma_{\rm meas}^2}\right) = 1 + \frac{\bmath g^{\rm T}
\mathbfss Q^{-1} \bmath g}{\sigma_{\rm meas}^2},
\label{.}
\end{equation}
which directly implies~(\ref{J-simple}). Also, we can use the identity $(\mathbfss A +
\bmath x \bmath y^{\rm T})^{-1} = \mathbfss A^{-1} - \mathbfss A^{-1} \bmath x \bmath
y^{\rm T}\mathbfss A^{-1} /(1 + \bmath y^{\rm T} \mathbfss A^{-1} \bmath x)$ from
\citep[Appendix~1]{ErmZhig}, which is valid for arbitrary matrix $\mathbfss A$ and vectors
$\bmath x,\bmath y$ with matching dimensions, to obtain
\begin{equation}
 \tilde\mathbfss C = \left(\mathbfss Q + \frac{\bmath g \bmath g^{\rm T}}{\sigma_{\rm
 meas}^2}\right)^{-1} = \mathbfss Q^{-1} - \frac{\bmath c \bmath c^{\rm T}}{\sigma^2},
\label{..}
\end{equation}
where $\bmath c = \mathbfss Q^{-1}\bmath g$. The relation~(\ref{l-simple}) is a direct
consequence of the latter equality. Using~(\ref{..}), we can also calculate
\begin{equation}
\tilde\mathbfss K = \mathbfss K - \frac{\bmath a \bmath a^{\rm T}}{\sigma^2}, \qquad
\frac{\det\tilde\mathbfss K}{\det\mathbfss K} = 1 - \frac{\bmath a^{\rm T}\mathbfss K^{-1}
\bmath a}{\sigma^2},
\label{...}
\end{equation}
where $\bmath a = \mathbfss A \bmath g$ with matrix $\mathbfss A$ given
in~(\ref{cond-var}). The first of equations~(\ref{...}) implies the
relation~(\ref{l-general}), and the second one implies the relation~(\ref{J-general}).

When $\mathbfss K$ and $\tilde\mathbfss K$ are degenerated due to dependency of the
parameters $\boldeta$, we can easily check the equality
\begin{equation}
\frac{\mathcal D \mathbfss K}{\mathcal D \tilde\mathbfss K} = 1 - \frac{\bmath a^{\rm
T}\mathbfss K^+ \bmath a}{\sigma^2}
\end{equation}
using~(\ref{...}) and the eigenvalue decompositions of the matrices $\mathbfss K$ and
$\tilde\mathbfss K$, also bearing in mind that the orthogonal matrix, needed to transform
$\mathbfss K$ to a diagonal form, coincides with that for $\tilde\mathbfss K$.

\bsp

\label{lastpage}

\end{document}